\documentclass{article}
\usepackage{spconf,amsmath,epsfig}
\usepackage{algorithm}
\usepackage{algorithmic}
\usepackage{graphicx, subfigure}

\title{SRP-MS: A New Routing Protocol for\\ Delay Tolerant Wireless Sensor Networks}

\name{N. Javaid$^{\ddag}$, A. A. Khan$^{\ddag}$, M. Akbar$^{\ddag}$, Z. A. Khan$^{\$}$, U. Qasim$^{\sharp}$}

\address{$^{\ddag}$COMSATS Institute of Information Technology, Islamabad, Pakistan. \\
        $^{\$}$Faculty of Engineering, Dalhousie University, Halifax, Canada.\\
        $^{\sharp}$University of Alberta, Alberta, Canada.}

\begin{document}
%
\maketitle
\begin{abstract}
Sink Mobility is becoming popular due to excellent load balancing between nodes and ultimately resulting in prolonged network lifetime and throughput. A major challenge is to provide reliable and energy-efficient operations are to be taken into consideration for different mobility patterns of sink. Aim of this paper is lifetime maximization of Delay Tolerant Wireless Sensor Networks (WSNs) through the manipulation of Mobile Sink (MS) on different trajectories. We propose Square Routing Protocol with MS (SRP-MS) based on existing SEP (Stable Election Protocol) by making it Cluster Less (CL) and introducing sink mobility.
\end{abstract}
\section{Introduction}
\label{sec:intro}

WSNs are based on tiny sensor nodes which have limited energy. Wireless Sensors are being used in medical, environment monitoring, in security surveillance, etc. Strive for the betterment of WSNs is still motivating to work on different techniques and modifications to enhance the lifetime and low energy consumption.  The main constraint in WSNs is the lifetime due to limited energy of the sensors. Lot of work has been done and yet many doors are still open to explore. Purpose of this work is to maximize the lifetime of WSNs without affecting the cost of network. Main focus of the paper is on sink mobility and its motion on different trajectories. In \cite{1} speed of the moving sink is discussed, its fast or slow in both cases lifetime of the network will improve because mobility increases the dimension (degree of freedom) of the problem.

As sinks are proficient machines which are provided with sufficient energy (able to refill). Sensor which are deployed in the field act as sources, as they gather data and provide the needed information. Sensor nodes send their sensed data to the sink for further processing. The data which is transmitted to the sink is send either in pull model or push model. Sensors send their data actively to the sink in push model and in pull they will send data the sensed data only on the sinks request. As the sink is sometimes out of the range of many sensors and data sending takes much energy. If transmission is multi hop, the nodes which are located near sink deplete their batteries first as they will act as rely for the farthest sensors. Another drawback is that if every node near the sink remains in active mode then it creates a bottle neck near the sink and the connection of farthest nodes with the sink is disturbed.

To avoid it, we are introducing MS in our proposed scheme SRP. Its variants are use to find the shortest path between sensors and sink. When sink is moving on the predefined trajectory the sensor nodes in the field and gathering data, only sensors in the sensing range send their data to the sink. Other nodes which are out from sensing range go in sleep mode until sink arrives.

In this paper, we extend conventional routing protocol Stable Election Protocol (SEP) \cite{2} removing its clustering mechanism and introducing MS in the field. Then we enhance the data collection in WSNs by MS in clusteless network. Here, we implement three different mobility patterns in the field to efficiently gather data in our proposed protocol SRP.

The remainder of this paper is organized as follows. In Section 2, related research works including MS. Section 3, discusses our motivations. Section 4 gives our network model description and details our proposed protocol SRP for efficiently data gathering with MS in WSN environments. Simulation results have been discussed in section 5. Finally, concluding remarks are presented in section 6.

\section{Related Work}
\label{sec:format}

Focus of the research in WSNs is on MS, as it is improving the lifetime of the network. Sink mobility can be consider in two categories, controlled (MS moves along pre-defined trajectory) \cite{3}, \cite{4} and uncontrolled (MS has random motion) \cite{5}. Random motion of sink if it stays on all the given set of location has polynomial solution for maximization of network life. However, controlling the sink motion in a specific trajectories is more challenging. Also, in \cite{6} mobile relay approach is discussed, in which MS receive data from nodes through direct transmissions and data transmission is with mechanical movement. Tolerable delay is introduced to avoid over flows and ques \cite{7}. In \cite{8} MS lowers the saturation from the nodes which are close to sink and resulting increase the network life. Authors in \cite{9} surveyed variants of Distributed Energy Efficient Clustering (DEEC) on basis of multi level heterogeneous network  to two level heterogeneous network. However, hierarchal clustering is discussed in \cite{10} by authors. They have used primary and secondary CHs in their clustering hierarchy. In \cite{11} the uniform distribution of nodes is used in planed region, hence take full advantage of three level of node heterogeneity.

Communication between sink and the wireless sensor nodes is perform through routing. Their are many techniques for routing, through clustering, multi-hop or direct. If a node has to send the data to sink which is very far then maximum energy is consume in transmission. To save energy and prolong lifetime of the network clustered base techniques are used in which nodes and sink both are stationary. In \cite{12} introduced a hierarchical clustering algorithm for sensor networks known as Low Energy Adaptive Clustering Hierarchy (LEACH). LEACH is clustered based routing protocol and cluster formation is distributed. CH election is random, and it rotates this clustering process, to efficiently distribute energy among sensor nodes. However, authors used clustering to evenly balance load between sensor nodes in \cite{2}, every node in two level heterogeneous hierarchal network selects itself to be a CH on bases of its initial energy compare to other nodes. Authors in \cite{13}, used clustering with three level of node heterogeneity in terms of energy to prolong network lifetime and stability period. Whereas, clustering technique is used by authors to reduce energy consumption in \cite{14}, CHs are elected probabilistically on the basis of ratio between residual energy of each node and average energy of network. However, all above mentioned papers they have not discussed sink mobility for prolonging network lifetime.

\section{Motivation}
SEP, LEACH, Threshold sensitive Energy Efficient sensor Network protocol (TEEN) \cite{15}, and Distributed Energy Efficient Clustering (DEEC) \cite{14} worked on clustering technique. All sensor nodes send their data to CHs and then CHs forward data to the sink. As sink is static in sensing field, maximum energy is consumed in transmitting data to sink. In order to ensure energy-efficient and reliable communication, this work focuses on MS. We implement various patterns of sink mobility in squared network in our proposed protocol SRP. From residual energy aspect, sink is mobile and nodes switch off their transmitters when they get out of the range of sink. Hence, save energy and performance enhanced than other conventional routing protocols. Our current goal is to achieve a robust self-configured WSN that maximizes lifetime. In above mentioned schemes maximum energy of the sensors is consume in electing CH. Here, in our proposed scheme we introduced MS instead of clustering. Now, sink moves on predefined trajectory and collect data directly. Nodes after transmission go to sleep mode and sink moves on.


\section{SRP-MS: Proposed Scheme}
\label{sec:pagestyle}

There are $N$ number of sensor nodes deployed randomly in the area of $100\times100$ and sink is moving at predefined path within the sensing field. We propose SRP and its variants. We introduce sink mobility in Squared path within the Squared region (SS-SRP). We implement sink mobility in Circular path within the Squared field (SC-SRP) with three different radii. However, in third variant we implement sink mobility in Circular path within the Circular (CC-SRP) sensing field. Strive is to improve the lifetime of the network by introducing two types of nodes normal and advance, and making the sink mobile. In this model sink is mechanically driven and can be recharged, so energy is not a constraint on moving sink, in short we can say that sink as a small vehicle, which is unmanned, and transceiver is attached with it. Mobile sink collects data, not randomly but on the defined path. To avoid buffering over flow of the information packets received at nodes, the tour of the sink and its sojourn locations have specific time, so that all nodes in the network can easily transfer their data without any loss of packets. To make it cost-effective sojourn tour is predefined and the distance between the two locations is bounded by $r_{max}$. By exploiting the trajectories of sink, we explored different results. If sink moves in the circular pattern within the network it gives best results as compared to other squared mobility pattern. Observed that sum of the sojourn locations is actually the network lifetime.

\begin{figure}[ht] 
\hspace{-0.8cm}
\includegraphics [height=5 cm, width=10 cm]{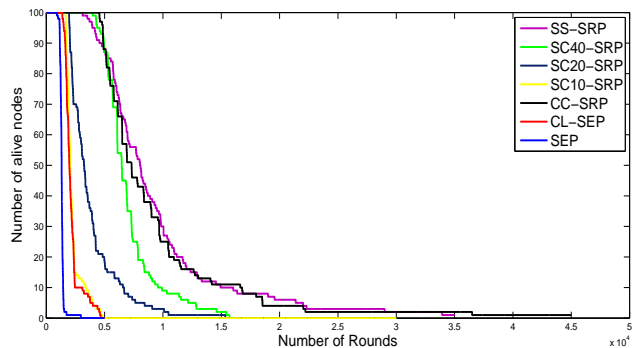}
\vspace{-0.2cm}
\caption{Alive nodes}
\end{figure}

\section{Simulation Experiments}
\label{sec:typestyle}

During simulations, we propose both square and circular fields with randomly (uniformly) deployed nodes. Following the patterns of MS, we discuss throughput and rates of the dead nodes. Applied a circular pattern of MS in the square field with different radii, $10m$, $20m$ and $40m$. Then square inside a square field and circle inside the circular field. Finally we compare these results with SEP \cite{2}, and CL-SEP. Network parameters are defined in table 1.

Comparing the graphs of variants of SRP-MS in SS-SRP and CC-SRP, the trajectory of SS-SRP is given in fig. $3$. First node of SS-SRP dies around $3000$ round and in CC-SRP at $4100$ round. Last node of CC-SRP dies at $45000$ rounds and in SS-SRP at $35000$ rounds. Dimensions of the square field are $100m\times100m$ MS trajectory is also square and its is moving along the perimeter of $50m \times 50m$ square which is exactly inside the WSN square field. Sensing range of a sink is $35.35m$ because of the far most nodes on the corner of the field. MS is gathering data efficiently in this way. The nodes placed near to the Square trajectory of sink are maximum time exposed to the sensing range of the sink and stay in awake mode for the longer time as compare to the nodes placed outside. 


\begin{figure}[ht] 
\centering
\hspace{0.4cm}
\includegraphics [height=5 cm, width=7 cm]{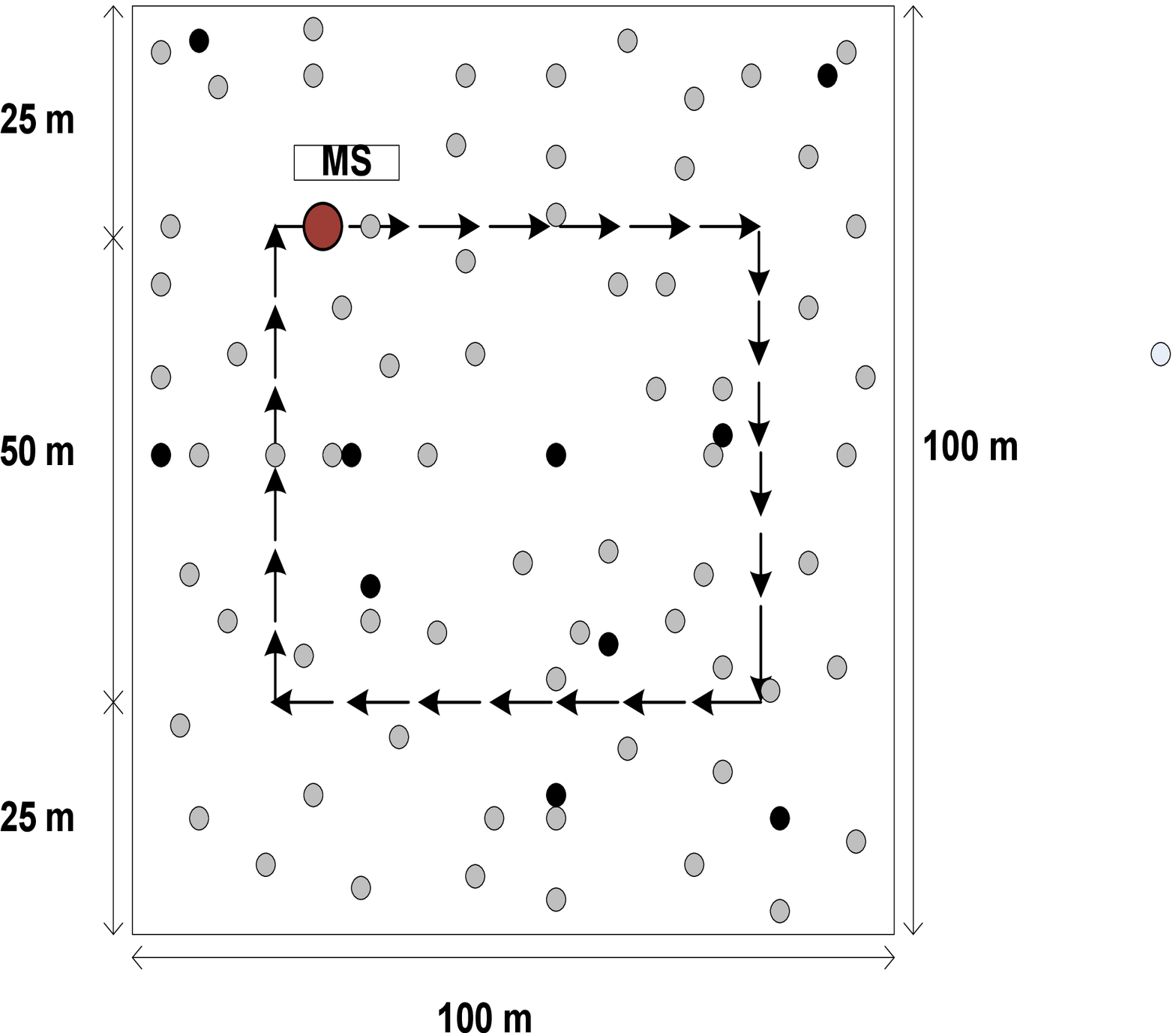}
\vspace{-0.2cm}
\caption{Squared sink mobility in squared field}
\end{figure}

 \begin{figure}[ht] 
\centering
\hspace{0.4cm}
\includegraphics [height=5 cm, width=6 cm]{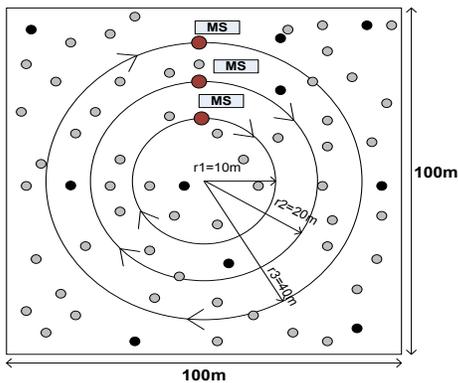}
\vspace{-0.2cm}
\caption{Circular sink mobility in squared field}
\end{figure}

Mobile sink collects data directly from sensor nodes by one hop communication known as direct contact data collection. Data may be retransmitted by the mobile sink if needed. This technique minimizes energy consumption between sensors for communication since sensors do not need to forward messages for each other. These two variants are performing well as compare to others because the MS is receiving maximum data due to the well balanced trajectories. Every sensor node in both fields is directly transmitting sensed data to MS. An other simulation experiment is done by using MS in a circular trajectory with in a square field. Three variants are compared in terms of alive nodes and results are shown in fig. $2$. and their trajectories are shown in fig. $3$. MS is varied in a circle with $3$ different radii $(40m, 20m, 10m)$. Observe that, death of first node of SC$40$-SRP lies in range of $3700th$ round. This technique beats SC$20$-SRP, SC$10$SRP, CL-SEP and SEP in stability period and network lifetime. This is because in SC$40$-SRP sink is moving at radius of $40$ at circular path inside the square field and having sensing range defined as to be $40$ because maximum distance from trajectory to corner of the field is $31.35$ and distance from trajectory to the center point of the field is $40$. So, even if sensor is at corner or at the center of the field would come in sensing range of mobile sink and whole network would be covered. However, if we talk about SC$20$-SRP it performs poor than SC$40$-SRP, because in this sink is moving at radius of $20$ in circular trajectory inside the square field with sensing range of $51.35$. So, nodes inside the trajectory always exposed to mobile sink because of always coming in the sensing range of mobile sink and drains their energy earlier. Nodes outer to the circular trajectory also feels maximum transmission distance when comes in sensing range hence consume more energy as compares to SC$40$-SRP.

\begin{table}[ht]
\centering
\begin{tabular}{|c |c |} 
\multicolumn{2}{c}{Table-1 Simulation parameters}\\
\hline
$ E_{elect}$ & 50nJ/bit\\
\hline                  
  $E_{DA}$ & 5nJ/bit/message  \\
\hline
 $\epsilon_{fs} $& 10pJ/bit/$m^2$  \\
\hline
 $\epsilon_{mp} $& 0.0013pJ/bit/$m^4$    \\
\hline
 $E_o$ & 0.5J \\
 \hline
 $K$ & 4000 \\
 \hline
 $P_{opt}$ & 0.1 \\
\hline
 $n$ & 100\\
 \hline
 $\alpha$ & 1 \\
 \hline
 $m$ & 0.1 \\
\hline
\end{tabular}
\end{table}

If we compare SC$10$-SRP with SC$20$-SRP whose topology is shown in fig. $3$. Simulation results in fig. $2$ shows that SC$20$-SRP beats SC$10$-SRP and in stability period and network lifetime because in SC$10$-SRP sink is moving in circular trajectory with radius $10$. Sensing range of the mobile sink here in this scenario is defined as $61.71$, as sensing range is very large and nodes within the circular trajectory remain alive and do not go in sleeping mode hence consume their available energy rapidly and die earlier.
 Hot spot problem arises in multi-hop communication with static sink. This results in making the network disconnected, though most of the sensors are still alive and working. From our simulations we can see that CL-SEP outperforms SEP, as SEP is clustered routing protocol and CL-SEP is clusterless routing protocol. CL-SEP performs better because there is no issue of relay nodes and nodes just do single transmission whereas in SEP nodes first send their data to CHs and then CHs send the aggregated data to sink. Hence consuming more energy in double transmission.

%

Throughput comparison of above discussed techniques, we can analyse that CC-SRP has highest throughput among all techniques.
 Whereas SS-SRP has second highest throughput in above discussed techniques its throughput increases up to $2.5\times10^5$ at $20,000$ rounds and then goes constant. However, among SC$40$-SRP, SC$20$-SRP and SC$10$-SRP, throughput of SC$40$-SRP has highest throughput.
  CL-SEP uses just single transmission whereas in conventional SEP a clustering technique is used in which the sensor nodes first send their data to associated CHs and then data is further forwarded to static sink via CHs hence consume more energy in double transmission.

\vspace{-.7cm}
\section{Conclusion}
\vspace{-.5cm}
\label{sec:majhead}

The use of MS in larger networks very useful in order to cover large areas and minimize the energy consumption at large transmission distances. In this paper we proposed energy efficient technique SRP with MS in squared (and circle) regions to prolong the network lifetime and stability period. Our approach uses different mobility patterns and compared the results in maximization of network life and stability period. Simulation results have shown that CC-SRP significantly prolongs the network life time and stability period when MS is moving in the circular path inside the circular field at an optimized radius.


\end{document}